# A MODIFIED VERSION OF THE INFERENCE FUNCTION FOR MARGINS AND INTERVAL ESTIMATION FOR THE BIVARIATE CLAYTON COPULA SUR TOBIT MODEL: A SIMULATION APPROACH

Paulo H. Ferreira[1] and Francisco Louzada[2]

[1] Department of Statistics
  Universidade Federal de São Carlos, Brazil
  phfs205@hotmail.com

[2] Institute of Mathematical Sciences and Computation
  Universidade de Sao Paulo, Brazil
  louzada@icmc.usp.br

**Abstract**

This paper extends the analysis of bivariate seemingly unrelated regression (SUR) Tobit model by modeling its nonlinear dependence structure through the Clayton copula. The ability in capturing/modeling the lower tail dependence of the SUR Tobit model where some data are censored (generally, at zero point) is an additionally useful feature of the Clayton copula. We propose a modified version of the inference function for margins (IFM) method (Joe and Xu, 1996), which we refer to as MIFM method, to obtain the estimates of the marginal parameters and a better (satisfactory) estimate of the copula association parameter. More specifically, we employ the data augmentation technique in the second stage of the IFM method to generate the censored observations (i.e. to obtain continuous marginal distributions, which ensures the uniqueness of the copula) and then estimate the dependence parameter. Resampling procedures (bootstrap methods) are also proposed for obtaining confidence intervals for the model parameters. A simulation study is performed in order to verify the behavior of the MIFM estimates (we focus on the copula parameter estimation) and the coverage probability of different confidence intervals in datasets with different percentages of censoring and degrees of dependence. The satisfactory results from the simulation (under certain conditions) and empirical study indicate the good performance of our proposed model and methods where they are applied to model the U.S. ready-to-eat breakfast cereals and fluid milk consumption data.

# 1. Introduction

As pointed out by Chen and Zhou (2011), researchers are often faced with the joint problem of censoring and simultaneity when working with microeconomic data. One class of models for which these issues arise is the multivariate Tobit models, which generalize univariate Tobit models to systems of equations. There are several generalizations available in the literature, and each of them is designed to capture several characteristics unique to each specific application (see Lee (1993) for a survey). In this paper we consider the seemingly unrelated regression (SUR) Tobit model, which is a SUR-type model where all limited dependent variables are partially observed or censored. See Zellner (1962), Greene (2003, p.341), and Zellner and Ando (2010) for more details on the SUR model, and Amemiya (1984) for a comprehensive review on the Tobit models.

Generally, the SUR Tobit model is not well applied due to the difficulty in the estimation. Various methods have been proposed to estimate the SUR Tobit model. See, for instance, Wales and Woodland (1983), Brown and Lankford (1992), and Kamakura and Wedel (2001) for the maximum likelihood (ML) estimation, Huang *et al.* (1987) for the expectation-maximization, Meng and Rubin (1996) for the expectation-conditional maximization (ECM), and Huang (1999) for the Monte Carlo ECM. However, all these methods are cumbersome, especially in high dimensions (Trivedi and Zimmer, 2005). Huang (2001), Taylor and Phaneuf (2009), and Baranchuk and Chib (2008) implement the SUR Tobit model through the Bayesian simulation-based methods, while Chen and Zhou (2011) consider semiparametric estimation of the model parameters. Nevertheless, all these estimation methods assume normal marginal error distributions. Additionally, modeling the dependence structure of the SUR Tobit model through the bivariate or multivariate normal distributions is limited to the linear relationship among marginal distributions by the correlation coefficients (Wichitaksorn *et al.*, 2012).

In order to relax the assumption on the linear dependence structure, mainly, we can apply copulas to analyze the SUR Tobit model (Wichitaksorn *et al.*, 2012). See, for instance, Joe (1997) and Nelsen (2006) for more details on copulas. The copulas have been well applied in several financial and economic applications with continuous and discrete margins (Pitt *et al.*, 2006; Smith and Khaled, 2012; and Panagiotelis *et al.*, 2012). However, the case of censored (or semi-continuous) margins has not been well studied and applied, as pointed out by Wichitaksorn *et al.* (2012). Additionally, the tail coefficients, especially the lower tail coefficient, from some copulas can reveal the dependence at the tails where some data are censored. But Trivedi and Zimmer (2005) implement the bivariate SUR Tobit model through some copulas (Clayton, Frank, Gaussian, and Farlie-Gumbel-Morgenstern copulas) to model the U.S. out-of-pocket and non-out-of-pocket medical expenses data and find that the two-stage ML/IFM estimation results are unstable. Yen and Lin (2008) estimate the copula-based censored equation system through the quasi-ML estimation method but consider only the Frank copula. Wichitaksorn *et al.* (2012) apply and combine the data augmentation techniques by Geweke (1991), Chib (1992), Chib and Greenberg (1998), Pitt *et al.* (2006), and Smith and Khaled (2012) to simulate the unobserved marginal dependent variables and proceed the bivariate SUR Tobit model implementation through the Bayesian Markov Chain Monte Carlo approach as in other copula models with continuous margins.

In this paper, we propose a modified version of the IFM method (Joe and Xu, 1996), hereafter MIFM method, to implement the bivariate SUR Tobit model through an Archimedean copula that appears regularly in statistics literature: the Clayton copula. Archimedean copulas, such as Clayton, Frank and Gumbel copulas, according to Trivedi and Zimmer (2005), are popular because they accommodate different patterns of dependence and have relatively straightforward functional forms. However, in this work the usage of Clayton copula becomes more attractive because of its ability in modeling the lower tail dependence of the SUR Tobit model where some data are censored (at zero point, here). In short, the MIFM method proposed here employ the data augmentation algorithm, presented in Wichitaksorn *et al.* (2012, Appendix A, Algorithm A2), in the second stage of the IFM method to generate the censored observations and obtain a better (satisfactory) estimate of the copula dependence parameter. The idea behind such a modification is to meet the Sklar's theorem (Sklar, 1959), which states that marginal distributions must be continuous to ensure the uniqueness of the resulting copula. Since the asymptotic approximation is problematic in this case, we employ resampling procedures (a parametric resampling plan) for obtaining confidence intervals for the model parameters. More specifically, we consider the percentile, bias corrected and accelerated (BCa) (Efron and Tibshirani, 1993), and basic (Davison and Hinkley, 1997) methods for constructing bootstrap confidence intervals.

This paper is organized as follows. Section 2 describes the bivariate SUR Tobit model and the Clayton copula. Section 3 shows the model implementation through the MIFM method and the confidence intervals construction from the bootstrap distributions of model parameters. Section 4 presents the simulation study. Section 5 provides the application. Finally, some final remarks in Section 6 conclude the paper.

## 2. The Bivariate Clayton Copula-Based SUR Tobit Model

The SUR Tobit model with two dependent variables, or simply bivariate SUR Tobit model, is expressed as

$$y_{ij}^* = x_{ij}^{'}\beta_j + \varepsilon_{ij},$$

where $y_{ij} = y_{ij}^*$ if $y_{ij}^* > 0$ or $y_{ij} = 0$ otherwise, for $i = 1,..., n$ and $j = 1, 2$, where $n$ is the number of observations, $y_{ij}$ is the observed dependent variable of margin $j$, $x_{ij}$ is the $k \times 1$ vector of covariates, $\beta_j$ is the $k \times 1$ vector of regression coefficients (estimable parameters), and $\varepsilon_{ij}$ is the margin $j$'s error that we assume follows a normal distribution with zero mean and variance $\sigma_j^2$ (this assumption about the error terms characterizes the standard Tobit model or Type I Tobit model, proposed by Amemiya, 1984).
Given these error terms, the density function of $y_{ij}$ is (Trivedi and Zimmer, 2005),

$$f_j\left(y_{ij} \mid x_{ij}, \beta_j, \sigma_j\right) = \prod_{y_{ij}=0}\left[1 - \Phi\left(\frac{x_{ij}^{'}\beta_j}{\sigma_j}\right)\right] \prod_{y_{ij}>0} \phi\left(\frac{y_{ij} - x_{ij}^{'}\beta_j}{\sigma_j}\right), \qquad (1)$$

where $\phi(.)$ and $\Phi(.)$ are the standard normal probability density function (p.d.f.) and cumulative distribution function (c.d.f.), respectively. The corresponding distribution function of $y_{ij}$, $F_j\left(y_{ij} \mid x_{ij}, \beta_j, \sigma_j\right)$, is obtained by replacing $\phi(.)$ with $\Phi(.)$ in the second part of (1).
Usually, the dependence between the error terms $\varepsilon_{i1}$ and $\varepsilon_{i2}$ is modeled via a bivariate distribution, especially the bivariate normal distribution. However, as commented before (in Section 1), one of the limitations in applying a bivariate distribution to the bivariate SUR Tobit model is the linear relationship between marginal distributions through the correlation coefficient. To overcome this problem, we can use copula function to model the nonlinear dependence structure in the bivariate SUR Tobit model with normal marginal distributions.

Thus, for the censored outcomes $y_{i1}$ and $y_{i2}$, the bivariate copula-based SUR Tobit distribution is given by

$$F(y_{i1}, y_{i2}) = C(u_1, u_2; \theta),$$

where $u_j = F_j(y_{ij} | \mathbf{x}_{ij}, \boldsymbol{\beta}_j, \sigma_j)$, for $j = 1, 2$, and $\theta$ is the copula association parameter, which is assumed to be a scalar.

In this paper we consider only the Clayton (1978) copula, which is also referred to as the Cook and Johnson (1981) copula and was originally studied by Kimeldorf and Sampson (1975). It takes the form

$$C(u_1, u_2; \theta) = \left(u_1^{-\theta} + u_2^{-\theta} - 1\right)^{-1/\theta}, \qquad (2)$$

with $\theta$ restricted on the region $(0,\infty)$. As the association parameter $\theta$ approaches zero, the marginals become independent. The Clayton copula does not allow negative dependence. As pointed out by Trivedi and Zimmer (2005), the Clayton copula is widely used to study correlated risks because it exhibits strong left tail dependence and relatively weak right tail dependence. In fact, when correlation between two events is strongest in the left tail of the joint distribution, Clayton is usually an appropriate modeling choice.

## 3. Inference

In this section, we discuss inference (point and interval estimation) for the parameters of the bivariate Clayton copula-based SUR Tobit model.

### 3.1. Estimation by the Modified Inference Function for Margins Method

Given the censored outcomes $y_{i1}$ and $y_{i2}$, $i = 1,..., n$, we can obtain the following log-likelihood function for the bivariate Clayton copula-based SUR Tobit model

$$l(\boldsymbol{\vartheta}) = \sum_{i=1}^{n} \log c\left(F_1(y_{i1} | \mathbf{x}_{i1}, \boldsymbol{\vartheta}_1), F_2(y_{i2} | \mathbf{x}_{i2}, \boldsymbol{\vartheta}_2); \theta\right) + \sum_{i=1}^{n} \sum_{j=1}^{2} \log f_j(y_{ij} | \mathbf{x}_{ij}, \boldsymbol{\vartheta}_j),$$

where $\boldsymbol{\vartheta} = (\boldsymbol{\vartheta}_1, \boldsymbol{\vartheta}_2, \theta)$ is the vector of model parameters, $\boldsymbol{\vartheta}_j = (\boldsymbol{\beta}_j, \sigma_j)$ is the margin $j$'s parameter vector, $f_j(y_{ij} | \boldsymbol{x}_{ij}, \boldsymbol{\vartheta}_j)$ is the p.d.f. of $y_{ij}$ (given by (1)), $F_j(y_{ij} | \boldsymbol{x}_{ij}, \boldsymbol{\vartheta}_j)$ is the c.d.f. of $y_{ij}$, and $c(u_1, u_2; \theta)$, with $u_j = F_j(y_{ij} | \boldsymbol{x}_{ij}, \boldsymbol{\vartheta}_j)$, is the p.d.f. of the Clayton copula, which is calculated from (2) as

$$c(u_1, u_2; \theta) = \frac{\partial^2 C(u_1, u_2; \theta)}{\partial u_1 \partial u_2} = (\theta + 1)(u_1 u_2)^{-(\theta+1)}(u_1^{-\theta} + u_2^{-\theta} - 1)^{-\left(\frac{2\theta+1}{\theta}\right)}.$$

For model estimation, the use of copula methods allows for the use of the two-stage ML method by Joe and Xu (1996), the inference functions for margins (IFM), which estimates the marginal parameters $\boldsymbol{\vartheta}_j$ in a first step by

$$\hat{\boldsymbol{\vartheta}}_{jIFM} = \arg\max_{\boldsymbol{\vartheta}_j} \sum_{i=1}^{n} \log f_j(y_{ij} | \boldsymbol{x}_{ij}, \boldsymbol{\vartheta}_j),$$

for $j = 1, 2$, and then estimates the association parameter $\theta$ given $\hat{\boldsymbol{\vartheta}}_{jIFM}$ by

$$\hat{\theta}_{IFM} = \arg\max_{\theta} \sum_{i=1}^{n} \log c\left(F_1(y_{i1} | \boldsymbol{x}_{i1}, \hat{\boldsymbol{\vartheta}}_{1IFM}), F_2(y_{i2} | \boldsymbol{x}_{i2}, \hat{\boldsymbol{\vartheta}}_{2IFM}); \theta\right).$$

Each maximization task has a small number of parameters, which reduces the shortness computing. However, the IFM method provides a biased estimate for the parameter $\theta$ due to the presence of censored observations in both marginals. Since we are interested in the bivariate Clayton copula-based SUR Tobit model where both marginal distributions are semi-continuous, we are dealing with the case that is not one-to-one relationship between the marginal distributions and the copula, i.e. there is more than one copula to join the marginal distributions. This constitutes a violation of the Sklar's theorem (Sklar, 1959). When it happens, researchers often face with problems in the model fitting and validation. To facilitate the implementation of copula model with semi-continuous margins, the semi-continuous marginal distributions can be augmented n order to achieve continuity. More specifically, we can employ the data augmentation technique to simulate the unobserved dependent variables in the censored margins, i.e. we generate the unobserved data with all properties, e.g., mean, variance, and dependence structure that match with the observed ones, and obtain a continuous marginal distributions (Wichitaksorn *et al.*, 2012). Thus, in order to obtain an unbiased estimate for the association parameter $\theta$, we replace $y_{ij}$ by the augmented data $y_{ij}^a$ in the second stage of the IFM method (MFIM method) and proceed the copula parameter estimation as usual in the case of continuous margins. See Wichitaksorn *et al.* (2012, Appendix A, Algorithm A2 for more details on one data augmentation algorithm we can employ). Such an algorithm is based on the conditional distribution of the Clayton copula and we can consider the estimates of the marginal parameters $\hat{\boldsymbol{\vartheta}}_{jIFM}$, $j = 1, 2$, and the copula dependence parameter estimated from the observed data in its implementation.

Finally, given the generated/augmented marginal dependent variable $y_{ij}^a$, we can estimate the association parameter $\theta$ by

$$\hat{\theta}_{MIFM} = \arg\max_{\theta} \sum_{i=1}^{n} \log c\left(F_1\left(y_{i1}^a \mid \boldsymbol{x}_{i1}, \hat{\boldsymbol{\vartheta}}_{1IFM}\right), F_2\left(y_{i2}^a \mid \boldsymbol{x}_{i2}, \hat{\boldsymbol{\vartheta}}_{2IFM}\right); \theta\right).$$

## 3.2. Interval Estimation

Joe and Xu (1996) combine the IFM method with the use of the jackknife method for estimation of the standard errors of the multivariate model parameter estimates. This eliminates the requirement for analytic derivatives to obtain the inverse Godambe information matrix or asymptotic covariance matrix associated with the vector of parameter estimates under some regularity conditions. However, as will be seen in Sections 4 and 5, the jackknife is not valid for obtaining standard errors of parameters estimates when using the MIFM method (the jackknife method produces an overestimate of the standard error of the association parameter estimator). This implies that confidence intervals for the bivariate Clayton copula-based SUR Tobit model parameters cannot be constructed using this method. To overcome this problem, we propose the usage of bootstrap methods (some of them do not require the calculation of the standard errors of the parameter estimators) for constructing confidence intervals.

Let $\vartheta_h$, $h = 1,..., k$, be a (any) component of the parameter vector $\boldsymbol{\vartheta}$ of the bivariate Clayton copula-based SUR Tobit model (see section 3.1). By using a resampling plan (non-parametric, parametric or semi-parametric resampling plan), we obtain the bootstrap estimates (via the MIFM method, for instance) $\hat{\vartheta}_{h1}^*, \hat{\vartheta}_{h2}^*, ..., \hat{\vartheta}_{hB}^*$ of $\vartheta_h$, where $B$ is the number of bootstrap samples (for 90-95 per cent confidence intervals, Efron and Tibshirani (1993) and Davison and Hinkley (1997) suggest that $B$ should be between 1000 and 2000). Then, we can derive confidence intervals from the bootstrap distribution, without using/calculating standard errors, by the following three methods, for instance:

- *Percentile bootstrap* (Efron and Tibshirani, 1993): the 100(1-2α)% percentile confidence interval is defined by the 100(α)th and 100(1-α)th percentiles of the bootstrap distribution of $\hat{\vartheta}_h^*$, i.e.

$$\left[ \hat{\vartheta}_h^{*(\alpha)} , \hat{\vartheta}_h^{*(1-\alpha)} \right]. \tag{3}$$

For Carpenter and Bithell (2000), simplicity is the attraction of this method. Moreover, no invalid parameter values can be included in the interval.

- *Bias-Corrected and Accelerated (BCa) bootstrap* (Efron, 1987; Efron and Tibshirani, 1993): the BCa interval endpoints are also given by percentiles of the bootstrap distribution, but the percentiles used depend on two numbers $\hat{a}$ and $\hat{z}_0$, called the *acceleration* and the *bias-correction*, respectively. The 100(1-2α)% BCa confidence interval is given by

$$\left[ \hat{\vartheta}_h^{*(\alpha_1)} , \hat{\vartheta}_h^{*(\alpha_2)} \right],$$

where

$$\alpha_1 = \Phi\left( \hat{z}_0 + \frac{\hat{z}_0 + z^{(\alpha)}}{1 - \hat{a}\left(\hat{z}_0 + z^{(\alpha)}\right)} \right) \quad \text{and} \quad \alpha_2 = \Phi\left( \hat{z}_0 + \frac{\hat{z}_0 + z^{(1-\alpha)}}{1 - \hat{a}\left(\hat{z}_0 + z^{(1-\alpha)}\right)} \right).$$

Here, $z^{(\alpha)}$ is the 100(α)th percentile point of a standard normal distribution. Note that if $\hat{a}$ and $\hat{z}_0$ are equal to zero, then the BCa interval is the same as the percentile interval (3).

The value of the bias-correction $\hat{z}_0$ is obtained directly from the proportion of bootstrap replications/estimates less than the original estimate (from the original data) $\hat{\vartheta}_h$, i.e.

$$\hat{z}_0 = \Phi^{-1}\left( \frac{\#\left\{\hat{\vartheta}_h^* < \hat{\vartheta}_h\right\}}{B} \right),$$

with $\Phi^{-1}(.)$ indicating the inverse function of a standard normal c.d.f.

One way to compute the acceleration $\hat{a}$ is in terms of the jackknife values of the statistic of interest. Let $\hat{\vartheta}_{h(i)}$ be the value of the estimate of the parameter $\vartheta_i$ obtained from the original data with the *i*th point (observation) deleted, and define $\hat{\vartheta}_{h(.)} = \sum_{i=1}^{n} \hat{\vartheta}_{h(i)} / n$. A simple expression for $\hat{a}$ is

$$\hat{a} = \frac{\sum_{i=1}^{n} \left( \hat{\vartheta}_{h(.)} - \hat{\vartheta}_{h(i)} \right)^3}{6 \left\{ \sum_{i=1}^{n} \left( \hat{\vartheta}_{h(.)} - \hat{\vartheta}_{h(i)} \right)^2 \right\}^{3/2}}.$$

The BCa interval generally has a smaller coverage error than the percentile interval (Efron and Tibshirani, 1993).

- *Basic bootstrap* (Davison and Hinkley, 1997): the basic bootstrap is one of the simplest schemes to construct confidence intervals. We proceed in a similar way to the percentile bootstrap, using the percentiles of the bootstrap distribution of $\hat{\vartheta}_h^*$, but with the following different formula (note the inversion of the left and right quantiles!)

$$\left[ 2\hat{\vartheta}_h - \hat{\vartheta}_h^{*(1-\alpha)} \, , \, 2\hat{\vartheta}_h - \hat{\vartheta}_h^{*(\alpha)} \right]. \qquad (4)$$

Note that if there is a parameter constraint (such as $\vartheta_h > 0$) the 100(1-2α)% basic confidence interval given by (4) may include invalid parameter values.

Besides the three methods described above, we can also construct confidence intervals by the usual asymptotic approach, which requires the calculation of the standard errors as follows.

- *Standard normal interval* (Efron and Tibshirani, 1993): since most statistics are asymptotically normally distributed, in large samples we can therefore use the standard error estimate, $\hat{se}_h$, along with the normal distribution, to produce a 100(1-2α)% confidence interval for $\vartheta_h$ based on the original estimate $\hat{\vartheta}_h$:

$$\left[ \hat{\vartheta}_h - z^{(1-\alpha)} \cdot \hat{se}_h, \; \hat{\vartheta}_h - z^{(\alpha)} \cdot \hat{se}_h \right].$$

In this case, $\hat{se}_h$ is the $h$th entry on the diagonal of the estimated (via bootstrap or jackknife method) asymptotic covariance matrix, say $\hat{\Sigma}$, of the parameter vector estimate $\hat{\vartheta}$ of the bivariate Clayton copula-based SUR Tobit model. The bootstrap-based covariance matrix estimate is given by

$$\hat{\Sigma}_{boot} = \frac{1}{B-1} \sum_{b=1}^{B} \left( \hat{\vartheta}_b^* - \bar{\hat{\vartheta}}^* \right) \left( \hat{\vartheta}_b^* - \bar{\hat{\vartheta}}^* \right)',$$

where $\hat{\vartheta}_b^*$, $b = 1,\ldots, B$, is the bootstrap estimate of $\vartheta$ and

$$\bar{\hat{\vartheta}}^* = \left( \frac{1}{B} \sum_{b=1}^{B} \hat{\vartheta}_{1b}^*, \; \frac{1}{B} \sum_{b=1}^{B} \hat{\vartheta}_{2b}^*, \mathrm{K}, \; \frac{1}{B} \sum_{b=1}^{B} \hat{\vartheta}_{kb}^* \right).$$

The delete-one jackknife-based covariance matrix estimate is given by

$$\hat{\Sigma}_{jack} = \sum_{i=1}^{n} \left( \hat{\vartheta}_{(i)} - \hat{\vartheta} \right) \left( \hat{\vartheta}_{(i)} - \hat{\vartheta} \right)'$$

(Joe and Xu, 1996), where $\hat{\vartheta}$ is the original estimate of $\vartheta$ and $\hat{\vartheta}_{(i)}$ is the estimate of $\vartheta$ with the $i$th observation deleted, $i = 1,\ldots, n$.

## 4. Simulation Study

In this section, we present results of the performed simulation study. The simulation study was conducted to verify the behavior of the MIFM estimates (with a focus on the copula association parameter estimate) and the coverage probability of different confidence intervals (constructed using the methods described in section 3.2) for the bivariate Clayton copula-based SUR Tobit model parameters. Here, we considered some circumstances that may arise in the development of bivariate SUR Tobit models, and which involve the sample size, the censoring percentage (i.e. the percentage of zero observations) in the margins and the dependence degree between them.

## 4.1. General Specifications

In the simulation study, we apply the Clayton copula to model the dependence structure of the bivariate SUR Tobit. We set the true value for the association parameter $\theta$ at 0.25, 1.2, 2, 5 and 10, which correspond to a Kendall's tau association measure[1] of 0.11, 0.375, 0.50, 0.71 and 0.83, respectively. For the Clayton copula data generation, see McNeil *et al.* (2005, p.224).

For $i = 1,..., n$, the model errors $\varepsilon_{i1}$ and $\varepsilon_{i2}$ are assumed to follow a normal distribution with $\varepsilon_{i1} \sim N(0,\sigma_1)$ and $\varepsilon_{i2} \sim N(0,\sigma_2)$, where $\sigma_1 = 1$ and $\sigma_2 = 2$ are the standard deviation parameters (or scale parameters) for margins 1 and 2, respectively. The covariates for margin 1 are $x_{i1,0} = 1$ and $x_{i1,1}$ is randomly simulated from a standard normal distribution. While the covariates for margin 2 are generated as $x_{i2,0} = 1$ and $x_{i2,1}$ is randomly simulated from $N(1,2)$. To ensure a percentage of censoring for both margins of approximately 5%, 15%, 25%, 40% and 50%, we assume the following true values for $\boldsymbol{\beta}_1$ and $\boldsymbol{\beta}_2$: $\boldsymbol{\beta}_1 = (2,1)$ and $\boldsymbol{\beta}_2 = (4,-0.5)$, $\boldsymbol{\beta}_1 = (1.5,1)$ and $\boldsymbol{\beta}_2 = (3,-0.5)$, $\boldsymbol{\beta}_1 = (1,1)$ and $\boldsymbol{\beta}_2 = (2,-0.5)$, $\boldsymbol{\beta}_1 = (0.5,1)$ and $\boldsymbol{\beta}_2 = (1,-0.5)$, and $\boldsymbol{\beta}_1 = (0.1,1)$ and $\boldsymbol{\beta}_2 = (0.25,-0.5)$, respectively. We then generated $M = 100$ datasets each with $n = 200$, 500 and 1000, but for purposes of convenience we only present here the results for $n = 1000$, since also the results of interest for the others sample sizes are quite similar (the only observed differences were a bias and mean squared error reduction with increasing sample size). For each dataset (original sample), we obtain 1000 bootstrap samples via a parametric resampling plan (parametric bootstrap approach), i.e. we fit a bivariate Clayton copula-based SUR Tobit model to each dataset and then generate a set of 1000 new datasets (the same size as the original dataset) from this estimated parametric model. The computing language is written in R version 3.0.1.

We assess the performance of our proposed model and methods through the coverage probabilities of the confidence intervals (the nominal value here is 0.90 or 90%), the Bias and the Mean Squared Error (MSE), in which the Bias and the MSE of each parameter $\gamma$ are given by

$$Bias = M^{-1}\sum_{r=1}^{M}(\hat{\gamma}_r - \gamma) \quad \text{and} \quad MSE = M^{-1}\sum_{r=1}^{M}(\hat{\gamma}_r - \gamma)^2,$$

---

[1] The Kendall's tau for Clayton copula is equal to $\tau = \theta/(\theta + 2)$ (McNiel *et al.*, 2005, p.222).

respectively, where $M = 100$ is the number of replications (original datasets) and $\hat{\gamma}_r$ is the estimated value of $\gamma$ at the $r$th replication.

## 4.2. Simulation Results

In this section, we present the main results obtained from the simulation study performed with samples (datasets) of different sizes (but we only exhibit here the results for $n = 1000$, because of the reasons mentioned above in Section 4.1), percentages of censoring in the margins and degrees of dependence between them, regarding the bivariate Clayton copula-based SUR Tobit model parameters estimated by the MIFM method. We also show the results relating to the estimated coverage probabilities of the 90% confidence intervals for the model parameters, obtained through bootstrap methods (percentile, BCa, basic and standard normal intervals) and also via the usual delete-one jackknife method (standard normal intervals).

Figure 1 illustrates the resulting augmented data from the Clayton copula. Given the fixed parameter values, Figure 1 shows the augmented data generated by the algorithm presented in Wichitaksorn *et al.* (2012, Appendix A, Algorithm A2). Note that we only augment the data for the second, third and fourth quadrants (or regions), where one or both the margins $y_{i1}$ and $y_{i2}$ are less than zero. Since the most difficult task is to recover the information on the dependence structure, it is clear from the Figure 1 that the employed data augmentation technique can recover well the information that has been censored.

The estimation results from Tables 1-5 show that our proposed model (bivariate Clayton copula-based SUR Tobit model) and estimation method (MIFM method) produce good (unbiased) estimates for the association parameter $\theta$ only when the percentage of censoring in the margins is low (less than 15% for $\theta = 0.25$, less than 25% for $\theta = 1.2$, and less than 40% for $\theta = 2$) and/or the degree of dependence between them is high ($\theta = 5$ and $\theta = 10$). In these situations, the parameter means are close to the true values with the reasonable standard deviations; the coverage probabilities of the confidence intervals obtained through the four bootstrap methods (with the exception of the percentile method and a few times the BCa method) are sufficiently high and close to the nominal value of 0.90, while the Biases and MSEs are relatively low. Notice that the usual jackknife method, proposed by Joe and Xu (1996), always produces confidence intervals for $\theta$ with coverage probability equal to 1. This occurs because the jackknife estimates of the standard error of the association parameter estimates are very high (if compared to the bootstrap estimates of the standard error, for instance), thus resulting in wide confidence intervals that always contain the true value of this parameter. Moreover, the confidence intervals based on jackknife method usually give negative lower limits, which does not make any sense here, since the Clayton copula parameter $\theta$ must be greater than zero). Overall, the basic bootstrap method, proposed by Davison and Hinkley (1997) performed well, independently of the degree of dependence between the margins and their percentages of censoring.

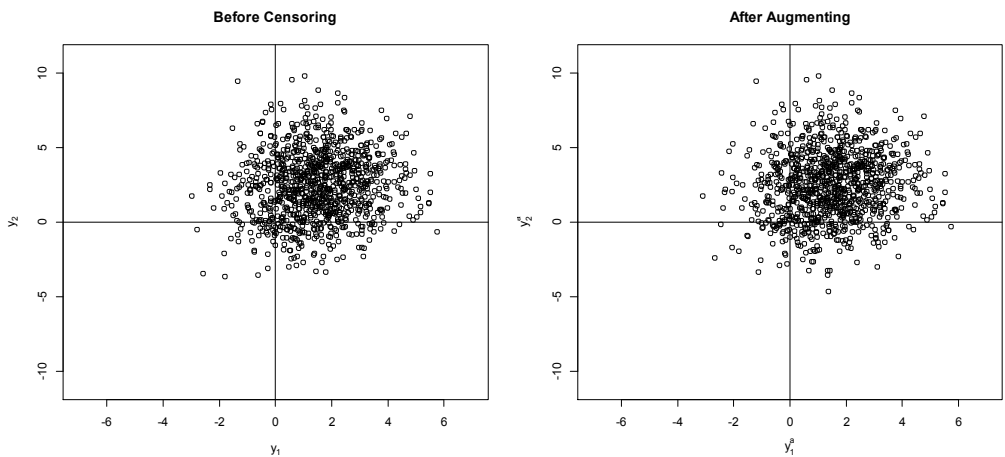

**Figure 1:** Augmented data from the bivariate Clayton copula-based SUR Tobit with normal errors, $\theta = 2$, $\boldsymbol{\beta}_1 = (1.5,1)$ and $\boldsymbol{\beta}_2 = (3,-0.5)$.

## 5. Case Study

In this section, we illustrate the applicability of our proposed model and methods to a part of a dataset extracted from the 1994-1996 Continuing Survey of Food Intakes by Individuals (CSFII) database (USDA, 2000). In the CSFII, two nonconsecutive days of dietary data for individuals of all ages residing in the United States were collected through in-person interviews using 24-hours recall. Each sample person reported the amount of each food item consumed. Where two days were reported there is also a third record containing daily averages. Socioeconomic and demographic data for the sample households and their members were also collected in the CSFII.

In this application, the relationship between the reported ready-to-eat breakfast cereals (RTEC) and fluid milk consumption (amount consumed, in 100 grams) of 500 adults (we only consider one member per household) age 20 or older is modeled by the bivariate SUR Tobit with normal margins through Clayton copula. This is the case where we expect the positive relationship between the consumption of RTEC and fluid milk. We include gender and age as the covariates and use them for both margins. From Figure 2 variable histograms, we find the normality assumption for the data seems to be reasonable, especially for the RTEC consumption data with Lilliefors (Kolmogorov-Smirnov) normality test p-value equal to 0.1691. Table 6 provides the definitions and sample statistics for all considered variables, where we observe the proportions of consuming individuals in the dataset to range from 90% for RTEC, to 91.4% for fluid milk. Among those consuming, an individual on average consumes 487 g of RTEC, and 2,828 g of fluid milk during the two-day period.

**Table 1:** Estimation results of bivariate Clayton copula-based SUR Tobit with normal margins ($\theta = 0.25$).

| | T.V. | Mean | S.D. | Bias | MSE | C.P. Standard Normal (Jackknife) | Standard Normal (Bootstrap) | Percentile | BCa | Basic |
|---|---|---|---|---|---|---|---|---|---|---|
| $\beta_{01}$ | 2 | 2.0056 | 0.0326 | 0.0056 | 0.0011 | 0.89 | 0.89 | 0.88 | 0.87 | 0.90 |
| $\beta_{11}$ | 1 | 0.9950 | 0.0308 | -0.0050 | 0.0010 | 0.92 | 0.92 | 0.93 | 0.92 | 0.92 |
| $\sigma_1$ | 1 | 0.9977 | 0.0218 | -0.0023 | 0.0005 | 0.94 | 0.95 | 0.92 | 0.94 | 0.94 |
| $\beta_{02}$ | 4 | 4.0037 | 0.0600 | 0.0037 | 0.0036 | 0.96 | 0.96 | 0.96 | 0.96 | 0.97 |
| $\beta_{12}$ | -0.5 | -0.5002 | 0.0341 | -0.0002 | 0.0012 | 0.91 | 0.91 | 0.91 | 0.89 | 0.91 |
| $\sigma_2$ | 2 | 1.9983 | 0.0461 | -0.0017 | 0.0021 | 0.89 | 0.89 | 0.89 | 0.89 | 0.89 |
| $\theta$ | 0.25 | 0.2575 | 0.0502 | 0.0075 | 0.0026 | 1.00 | 0.90 | 0.89 | 0.89 | 0.87 |
| $\beta_{01}$ | 1.5 | 1.5066 | 0.0329 | 0.0066 | 0.0011 | 0.91 | 0.91 | 0.90 | 0.91 | 0.90 |
| $\beta_{11}$ | 1 | 0.9935 | 0.0310 | -0.0065 | 0.0010 | 0.93 | 0.94 | 0.93 | 0.91 | 0.94 |
| $\sigma_1$ | 1 | 0.9968 | 0.0231 | -0.0032 | 0.0005 | 0.93 | 0.93 | 0.93 | 0.93 | 0.93 |
| $\beta_{02}$ | 3 | 3.0036 | 0.0606 | 0.0036 | 0.0037 | 0.97 | 0.96 | 0.97 | 0.97 | 0.96 |
| $\beta_{12}$ | -0.5 | -0.5004 | 0.0350 | -0.0004 | 0.0012 | 0.89 | 0.88 | 0.89 | 0.89 | 0.89 |
| $\sigma_2$ | 2 | 1.9986 | 0.0450 | -0.0014 | 0.0020 | 0.93 | 0.94 | 0.93 | 0.93 | 0.93 |
| $\theta$ | 0.25 | 0.2877 | 0.0537 | 0.0377 | 0.0043 | 1.00 | 0.87 | 0.63 | 0.88 | 0.94 |
| $\beta_{01}$ | 1 | 1.0043 | 0.0361 | 0.0043 | 0.0013 | 0.90 | 0.90 | 0.91 | 0.89 | 0.91 |
| $\beta_{11}$ | 1 | 0.9964 | 0.0331 | -0.0036 | 0.0011 | 0.91 | 0.91 | 0.91 | 0.91 | 0.91 |
| $\sigma_1$ | 1 | 0.9985 | 0.0276 | -0.0015 | 0.0008 | 0.90 | 0.90 | 0.89 | 0.88 | 0.87 |
| $\beta_{02}$ | 2 | 2.0044 | 0.0606 | 0.0044 | 0.0037 | 0.97 | 0.97 | 0.97 | 0.97 | 0.97 |
| $\beta_{12}$ | -0.5 | -0.5023 | 0.0353 | -0.0023 | 0.0013 | 0.90 | 0.91 | 0.90 | 0.91 | 0.91 |
| $\sigma_2$ | 2 | 1.9983 | 0.0516 | -0.0017 | 0.0027 | 0.92 | 0.91 | 0.91 | 0.91 | 0.92 |
| $\theta$ | 0.25 | 0.4083 | 0.0655 | 0.1583 | 0.0293 | 1.00 | 0.32 | 0.00 | 0.13 | 0.97 |
| $\beta_{01}$ | 0.5 | 0.5037 | 0.0425 | 0.0037 | 0.0018 | 0.85 | 0.84 | 0.85 | 0.84 | 0.86 |
| $\beta_{11}$ | 1 | 0.9968 | 0.0403 | -0.0032 | 0.0016 | 0.91 | 0.91 | 0.92 | 0.91 | 0.93 |
| $\sigma_1$ | 1 | 0.9988 | 0.0318 | -0.0012 | 0.0010 | 0.88 | 0.89 | 0.89 | 0.88 | 0.88 |
| $\beta_{02}$ | 1 | 1.0037 | 0.0647 | 0.0037 | 0.0042 | 0.96 | 0.96 | 0.96 | 0.95 | 0.96 |
| $\beta_{12}$ | -0.5 | -0.4992 | 0.0386 | 0.0008 | 0.0015 | 0.87 | 0.88 | 0.86 | 0.86 | 0.88 |
| $\sigma_2$ | 2 | 1.9977 | 0.0660 | -0.0023 | 0.0044 | 0.88 | 0.88 | 0.88 | 0.88 | 0.86 |
| $\theta$ | 0.25 | 0.6938 | 0.1098 | 0.4438 | 0.2090 | 1.00 | 0.02 | 0.00 | 0.00 | 0.90 |
| $\beta_{01}$ | 0.1 | 0.1008 | 0.0488 | 0.0008 | 0.0024 | 0.87 | 0.87 | 0.86 | 0.86 | 0.85 |
| $\beta_{11}$ | 1 | 0.9993 | 0.0459 | -0.0007 | 0.0021 | 0.88 | 0.87 | 0.87 | 0.86 | 0.85 |
| $\sigma_1$ | 1 | 1.0002 | 0.0339 | 0.0002 | 0.0011 | 0.87 | 0.88 | 0.88 | 0.86 | 0.88 |
| $\beta_{02}$ | 0.25 | 0.2476 | 0.0673 | -0.0024 | 0.0045 | 0.97 | 0.97 | 0.97 | 0.97 | 0.97 |
| $\beta_{12}$ | -0.5 | -0.4997 | 0.0421 | 0.0003 | 0.0018 | 0.88 | 0.87 | 0.86 | 0.86 | 0.86 |
| $\sigma_2$ | 2 | 2.0042 | 0.0678 | 0.0042 | 0.0046 | 0.94 | 0.94 | 0.94 | 0.94 | 0.94 |
| $\theta$ | 0.25 | 1.1322 | 0.1495 | 0.8822 | 0.8006 | 1.00 | 0.00 | 0.00 | 0.00 | 1.00 |

Notes: T.V. = True Value, S.D. = Standard Deviation, MSE = Mean Squared Error, and C.P. = Coverage Probability

**Table 2:** Estimation results of bivariate Clayton copula-based SUR Tobit with normal margins ($\theta = 1.2$).

| | | | | | | C.P. | | | | |
|---|---|---|---|---|---|---|---|---|---|---|
| | T.V. | Mean | S.D. | Bias | MSE | Standard Normal (Jackknife) | Standard Normal (Bootstrap) | Percentile | BCa | Basic |
| $\beta_{01}$ | 2 | 2.0056 | 0.0326 | 0.0056 | 0.0011 | 0.89 | 0.89 | 0.89 | 0.89 | 0.89 |
| $\beta_{11}$ | 1 | 0.9950 | 0.0308 | -0.0050 | 0.0010 | 0.92 | 0.92 | 0.92 | 0.92 | 0.91 |
| $\sigma_1$ | 1 | 0.9977 | 0.0218 | -0.0023 | 0.0005 | 0.94 | 0.94 | 0.92 | 0.95 | 0.94 |
| $\beta_{02}$ | 4 | 4.0087 | 0.0661 | 0.0087 | 0.0044 | 0.92 | 0.92 | 0.93 | 0.92 | 0.92 |
| $\beta_{12}$ | -0.5 | -0.5017 | 0.0329 | -0.0017 | 0.0011 | 0.92 | 0.91 | 0.92 | 0.92 | 0.91 |
| $\sigma_2$ | 2 | 1.9965 | 0.0405 | -0.0035 | 0.0016 | 0.96 | 0.96 | 0.96 | 0.97 | 0.96 |
| $\theta$ | 1.2 | 1.1980 | 0.0888 | -0.0020 | 0.0078 | 1.00 | 0.93 | 0.92 | 0.93 | 0.93 |
| $\beta_{01}$ | 1.5 | 1.5066 | 0.0329 | 0.0066 | 0.0011 | 0.91 | 0.90 | 0.90 | 0.91 | 0.89 |
| $\beta_{11}$ | 1 | 0.9935 | 0.0310 | -0.0065 | 0.0010 | 0.93 | 0.93 | 0.92 | 0.92 | 0.93 |
| $\sigma_1$ | 1 | 0.9968 | 0.0231 | -0.0032 | 0.0005 | 0.93 | 0.93 | 0.93 | 0.93 | 0.93 |
| $\beta_{02}$ | 3 | 3.0089 | 0.0671 | 0.0089 | 0.0045 | 0.92 | 0.93 | 0.94 | 0.94 | 0.93 |
| $\beta_{12}$ | -0.5 | -0.5023 | 0.0334 | -0.0023 | 0.0011 | 0.91 | 0.91 | 0.91 | 0.91 | 0.91 |
| $\sigma_2$ | 2 | 1.9967 | 0.0418 | -0.0033 | 0.0017 | 0.95 | 0.95 | 0.95 | 0.95 | 0.95 |
| $\theta$ | 1.2 | 1.2163 | 0.1041 | 0.0163 | 0.0110 | 1.00 | 0.91 | 0.90 | 0.90 | 0.91 |
| $\beta_{01}$ | 1 | 1.0043 | 0.0361 | 0.0043 | 0.0013 | 0.90 | 0.89 | 0.89 | 0.89 | 0.89 |
| $\beta_{11}$ | 1 | 0.9964 | 0.0331 | -0.0036 | 0.0011 | 0.91 | 0.90 | 0.90 | 0.91 | 0.90 |
| $\sigma_1$ | 1 | 0.9985 | 0.0276 | -0.0015 | 0.0008 | 0.90 | 0.90 | 0.89 | 0.89 | 0.90 |
| $\beta_{02}$ | 2 | 2.0082 | 0.0699 | 0.0082 | 0.0049 | 0.93 | 0.94 | 0.94 | 0.94 | 0.93 |
| $\beta_{12}$ | -0.5 | -0.5019 | 0.0344 | -0.0019 | 0.0012 | 0.92 | 0.92 | 0.92 | 0.92 | 0.92 |
| $\sigma_2$ | 2 | 1.9971 | 0.0511 | -0.0029 | 0.0026 | 0.90 | 0.91 | 0.90 | 0.91 | 0.91 |
| $\theta$ | 1.2 | 1.3119 | 0.1154 | 0.1119 | 0.0257 | 1.00 | 0.82 | 0.44 | 0.85 | 0.93 |
| $\beta_{01}$ | 0.5 | 0.5037 | 0.0425 | 0.0037 | 0.0018 | 0.85 | 0.85 | 0.84 | 0.85 | 0.85 |
| $\beta_{11}$ | 1 | 0.9968 | 0.0403 | -0.0032 | 0.0016 | 0.91 | 0.91 | 0.91 | 0.91 | 0.92 |
| $\sigma_1$ | 1 | 0.9988 | 0.0318 | -0.0012 | 0.0010 | 0.88 | 0.88 | 0.87 | 0.87 | 0.86 |
| $\beta_{02}$ | 1 | 1.0087 | 0.0696 | 0.0087 | 0.0049 | 0.94 | 0.94 | 0.96 | 0.95 | 0.94 |
| $\beta_{12}$ | -0.5 | -0.5016 | 0.0369 | -0.0016 | 0.0014 | 0.88 | 0.89 | 0.89 | 0.88 | 0.89 |
| $\sigma_2$ | 2 | 1.9961 | 0.0590 | -0.0039 | 0.0035 | 0.92 | 0.89 | 0.90 | 0.91 | 0.89 |
| $\theta$ | 1.2 | 1.5751 | 0.1662 | 0.3751 | 0.1681 | 1.00 | 0.25 | 0.00 | 0.03 | 0.90 |
| $\beta_{01}$ | 0.1 | 0.1008 | 0.0488 | 0.0008 | 0.0024 | 0.87 | 0.87 | 0.87 | 0.87 | 0.87 |
| $\beta_{11}$ | 1 | 0.9993 | 0.0459 | -0.0007 | 0.0021 | 0.88 | 0.87 | 0.86 | 0.86 | 0.87 |
| $\sigma_1$ | 1 | 1.0002 | 0.0339 | 0.0002 | 0.0011 | 0.87 | 0.88 | 0.86 | 0.86 | 0.88 |
| $\beta_{02}$ | 0.25 | 0.2562 | 0.0734 | 0.0062 | 0.0054 | 0.96 | 0.96 | 0.95 | 0.95 | 0.96 |
| $\beta_{12}$ | -0.5 | -0.5028 | 0.0423 | -0.0028 | 0.0018 | 0.88 | 0.89 | 0.88 | 0.88 | 0.87 |
| $\sigma_2$ | 2 | 1.9987 | 0.0689 | -0.0013 | 0.0047 | 0.94 | 0.94 | 0.93 | 0.93 | 0.94 |
| $\theta$ | 1.2 | 2.0221 | 0.2187 | 0.8221 | 0.7233 | 1.00 | 0.00 | 0.00 | 0.00 | 0.97 |

Notes: T.V. = True Value, S.D. = Standard Deviation, MSE = Mean Squared Error, and C.P. = Coverage Probability

**Table 3:** Estimation results of bivariate Clayton copula-based SUR Tobit with normal margins ($\theta = 2$).

| | | | | | | C.P. | | | | |
|---|---|---|---|---|---|---|---|---|---|---|
| | T.V. | Mean | S.D. | Bias | MSE | Standard Normal (Jackknife) | Standard Normal (Bootstrap) | Percentile | BCa | Basic |
| $\beta_{01}$ | 2 | 2.0056 | 0.0326 | 0.0056 | 0.0011 | 0.89 | 0.88 | 0.89 | 0.88 | 0.90 |
| $\beta_{11}$ | 1 | 0.9950 | 0.0308 | -0.0050 | 0.0010 | 0.92 | 0.92 | 0.92 | 0.92 | 0.91 |
| $\sigma_1$ | 1 | 0.9977 | 0.0218 | -0.0023 | 0.0005 | 0.94 | 0.94 | 0.93 | 0.94 | 0.94 |
| $\beta_{02}$ | 4 | 4.0103 | 0.0705 | 0.0103 | 0.0050 | 0.92 | 0.92 | 0.92 | 0.91 | 0.93 |
| $\beta_{12}$ | -0.5 | -0.5023 | 0.0321 | -0.0023 | 0.0010 | 0.93 | 0.92 | 0.92 | 0.92 | 0.92 |
| $\sigma_2$ | 2 | 1.9969 | 0.0407 | -0.0031 | 0.0017 | 0.96 | 0.96 | 0.95 | 0.97 | 0.96 |
| $\theta$ | 2 | 1.9859 | 0.1308 | -0.0141 | 0.0171 | 1.00 | 0.89 | 0.90 | 0.90 | 0.89 |
| $\beta_{01}$ | 1.5 | 1.5066 | 0.0329 | 0.0066 | 0.0011 | 0.91 | 0.91 | 0.88 | 0.88 | 0.90 |
| $\beta_{11}$ | 1 | 0.9935 | 0.0310 | -0.0065 | 0.0010 | 0.93 | 0.93 | 0.91 | 0.91 | 0.94 |
| $\sigma_1$ | 1 | 0.9968 | 0.0231 | -0.0032 | 0.0005 | 0.93 | 0.93 | 0.93 | 0.93 | 0.93 |
| $\beta_{02}$ | 3 | 3.0099 | 0.0713 | 0.0099 | 0.0051 | 0.92 | 0.93 | 0.92 | 0.91 | 0.93 |
| $\beta_{12}$ | -0.5 | -0.5031 | 0.0332 | -0.0031 | 0.0011 | 0.92 | 0.91 | 0.91 | 0.92 | 0.91 |
| $\sigma_2$ | 2 | 1.9981 | 0.0419 | -0.0019 | 0.0017 | 0.97 | 0.97 | 0.96 | 0.97 | 0.97 |
| $\theta$ | 2 | 2.0058 | 0.1282 | 0.0058 | 0.0163 | 1.00 | 0.93 | 0.93 | 0.94 | 0.96 |
| $\beta_{01}$ | 1 | 1.0043 | 0.0361 | 0.0043 | 0.0013 | 0.90 | 0.90 | 0.89 | 0.90 | 0.89 |
| $\beta_{11}$ | 1 | 0.9964 | 0.0331 | -0.0036 | 0.0011 | 0.91 | 0.91 | 0.91 | 0.90 | 0.91 |
| $\sigma_1$ | 1 | 0.9985 | 0.0276 | -0.0015 | 0.0008 | 0.90 | 0.89 | 0.89 | 0.88 | 0.88 |
| $\beta_{02}$ | 2 | 2.0104 | 0.0736 | 0.0134 | 0.0055 | 0.92 | 0.92 | 0.92 | 0.92 | 0.92 |
| $\beta_{12}$ | -0.5 | -0.5023 | 0.0336 | -0.0023 | 0.0011 | 0.91 | 0.91 | 0.91 | 0.91 | 0.91 |
| $\sigma_2$ | 2 | 1.9966 | 0.0492 | -0.0034 | 0.0024 | 0.92 | 0.92 | 0.91 | 0.92 | 0.92 |
| $\theta$ | 2 | 2.0620 | 0.1548 | 0.0620 | 0.0276 | 1.00 | 0.92 | 0.78 | 0.89 | 0.93 |
| $\beta_{01}$ | 0.5 | 0.5037 | 0.0425 | 0.0037 | 0.0018 | 0.85 | 0.86 | 0.85 | 0.85 | 0.84 |
| $\beta_{11}$ | 1 | 0.9968 | 0.0403 | -0.0032 | 0.0016 | 0.91 | 0.90 | 0.91 | 0.92 | 0.90 |
| $\sigma_1$ | 1 | 0.9988 | 0.0318 | -0.0012 | 0.0010 | 0.88 | 0.90 | 0.88 | 0.87 | 0.87 |
| $\beta_{02}$ | 1 | 1.0109 | 0.0738 | 0.0109 | 0.0055 | 0.94 | 0.92 | 0.91 | 0.92 | 0.94 |
| $\beta_{12}$ | -0.5 | -0.5033 | 0.0368 | -0.0033 | 0.0014 | 0.87 | 0.87 | 0.87 | 0.87 | 0.87 |
| $\sigma_2$ | 2 | 1.9963 | 0.0565 | -0.0037 | 0.0032 | 0.95 | 0.94 | 0.94 | 0.95 | 0.94 |
| $\theta$ | 2 | 2.2913 | 0.2106 | 0.2913 | 0.1288 | 1.00 | 0.69 | 0.08 | 0.65 | 0.90 |
| $\beta_{01}$ | 0.1 | 0.1008 | 0.0488 | 0.0008 | 0.0024 | 0.87 | 0.87 | 0.86 | 0.87 | 0.87 |
| $\beta_{11}$ | 1 | 0.9993 | 0.0459 | -0.0007 | 0.0021 | 0.88 | 0.87 | 0.86 | 0.87 | 0.87 |
| $\sigma_1$ | 1 | 1.0002 | 0.0339 | 0.0002 | 0.0011 | 0.87 | 0.87 | 0.83 | 0.88 | 0.87 |
| $\beta_{02}$ | 0.25 | 0.2617 | 0.0786 | 0.0117 | 0.0063 | 0.93 | 0.93 | 0.92 | 0.94 | 0.92 |
| $\beta_{12}$ | -0.5 | -0.5024 | 0.0413 | -0.0024 | 0.0017 | 0.88 | 0.87 | 0.90 | 0.89 | 0.86 |
| $\sigma_2$ | 2 | 1.9951 | 0.0688 | -0.0049 | 0.0047 | 0.94 | 0.95 | 0.95 | 0.95 | 0.95 |
| $\theta$ | 2 | 2.7131 | 0.2800 | 0.7131 | 0.5862 | 1.00 | 0.20 | 0.00 | 0.01 | 0.93 |

Notes: T.V. = True Value, S.D. = Standard Deviation, MSE = Mean Squared Error, and C.P. = Coverage Probability

**Table 4:** Estimation results of bivariate Clayton copula-based SUR Tobit with normal margins ($\theta = 5$).

|  | T.V. | Mean | S.D. | Bias | MSE | C.P. Standard Normal (Jackknife) | Standard Normal (Bootstrap) | Percentile | BCa | Basic |
|---|---|---|---|---|---|---|---|---|---|---|
| $\beta_{01}$ | 2 | 2.0056 | 0.0326 | 0.0056 | 0.0011 | 0.89 | 0.89 | 0.88 | 0.86 | 0.88 |
| $\beta_{11}$ | 1 | 0.9950 | 0.0308 | -0.0050 | 0.0010 | 0.92 | 0.92 | 0.92 | 0.92 | 0.92 |
| $\sigma_1$ | 1 | 0.9977 | 0.0218 | -0.0023 | 0.0005 | 0.94 | 0.94 | 0.93 | 0.95 | 0.93 |
| $\beta_{02}$ | 4 | 4.0132 | 0.0754 | 0.0132 | 0.0058 | 0.88 | 0.88 | 0.87 | 0.87 | 0.87 |
| $\beta_{12}$ | -0.5 | -0.5042 | 0.0311 | -0.0042 | 0.0010 | 0.91 | 0.90 | 0.88 | 0.88 | 0.91 |
| $\sigma_2$ | 2 | 1.9981 | 0.0394 | -0.0019 | 0.0015 | 0.97 | 0.97 | 0.97 | 0.96 | 0.96 |
| $\theta$ | 5 | 4.9621 | 0.2629 | -0.0379 | 0.0699 | 1.00 | 0.93 | 0.92 | 0.90 | 0.92 |
| $\beta_{01}$ | 1.5 | 1.5066 | 0.0329 | 0.0066 | 0.0011 | 0.91 | 0.91 | 0.89 | 0.88 | 0.90 |
| $\beta_{11}$ | 1 | 0.9935 | 0.0310 | -0.0065 | 0.0010 | 0.93 | 0.94 | 0.93 | 0.94 | 0.94 |
| $\sigma_1$ | 1 | 0.9968 | 0.0231 | -0.0032 | 0.0005 | 0.93 | 0.93 | 0.93 | 0.93 | 0.93 |
| $\beta_{02}$ | 3 | 3.0136 | 0.0761 | 0.0136 | 0.0059 | 0.89 | 0.89 | 0.89 | 0.89 | 0.89 |
| $\beta_{12}$ | -0.5 | -0.5042 | 0.0320 | -0.0042 | 0.0010 | 0.92 | 0.89 | 0.90 | 0.89 | 0.89 |
| $\sigma_2$ | 2 | 1.9973 | 0.0418 | -0.0027 | 0.0017 | 0.96 | 0.96 | 0.96 | 0.96 | 0.96 |
| $\theta$ | 5 | 4.9569 | 0.2595 | -0.0431 | 0.0685 | 1.00 | 0.95 | 0.95 | 0.94 | 0.94 |
| $\beta_{01}$ | 1 | 1.0043 | 0.0361 | 0.0043 | 0.0013 | 0.90 | 0.90 | 0.89 | 0.89 | 0.90 |
| $\beta_{11}$ | 1 | 0.9964 | 0.0331 | -0.0036 | 0.0011 | 0.91 | 0.91 | 0.91 | 0.91 | 0.91 |
| $\sigma_1$ | 1 | 0.9985 | 0.0276 | -0.0015 | 0.0008 | 0.90 | 0.90 | 0.88 | 0.86 | 0.90 |
| $\beta_{02}$ | 2 | 2.0132 | 0.0773 | 0.0132 | 0.0061 | 0.89 | 0.89 | 0.88 | 0.88 | 0.89 |
| $\beta_{12}$ | -0.5 | -0.5052 | 0.0319 | -0.0052 | 0.0010 | 0.94 | 0.94 | 0.92 | 0.91 | 0.93 |
| $\sigma_2$ | 2 | 1.9987 | 0.0417 | -0.0013 | 0.0017 | 0.97 | 0.97 | 0.97 | 0.97 | 0.97 |
| $\theta$ | 5 | 4.9822 | 0.2861 | -0.0178 | 0.0813 | 1.00 | 0.94 | 0.95 | 0.91 | 0.93 |
| $\beta_{01}$ | 0.5 | 0.4954 | 0.0372 | -0.0046 | 0.0014 | 0.91 | 0.92 | 0.90 | 0.91 | 0.92 |
| $\beta_{11}$ | 1 | 1.0056 | 0.0356 | 0.0056 | 0.0013 | 0.94 | 0.95 | 0.94 | 0.93 | 0.94 |
| $\sigma_1$ | 1 | 0.9977 | 0.0277 | -0.0023 | 0.0008 | 0.93 | 0.93 | 0.93 | 0.92 | 0.92 |
| $\beta_{02}$ | 1 | 0.9906 | 0.0771 | -0.0094 | 0.0060 | 0.90 | 0.90 | 0.90 | 0.90 | 0.90 |
| $\beta_{12}$ | -0.5 | -0.4993 | 0.0396 | 0.0007 | 0.0016 | 0.84 | 0.84 | 0.82 | 0.85 | 0.85 |
| $\sigma_2$ | 2 | 1.9999 | 0.0606 | -0.0001 | 0.0036 | 0.91 | 0.90 | 0.90 | 0.92 | 0.91 |
| $\theta$ | 5 | 5.1787 | 0.3801 | 0.1787 | 0.1749 | 1.00 | 0.91 | 0.80 | 0.90 | 0.89 |
| $\beta_{01}$ | 0.1 | 0.1008 | 0.0488 | 0.0008 | 0.0024 | 0.87 | 0.87 | 0.86 | 0.86 | 0.86 |
| $\beta_{11}$ | 1 | 0.9993 | 0.0459 | -0.0007 | 0.0021 | 0.88 | 0.88 | 0.88 | 0.87 | 0.88 |
| $\sigma_1$ | 1 | 1.0002 | 0.0339 | 0.0002 | 0.0011 | 0.87 | 0.87 | 0.85 | 0.86 | 0.88 |
| $\beta_{02}$ | 0.25 | 0.2627 | 0.0837 | 0.0127 | 0.0071 | 0.94 | 0.92 | 0.93 | 0.92 | 0.92 |
| $\beta_{12}$ | -0.5 | -0.5039 | 0.0405 | -0.0039 | 0.0016 | 0.90 | 0.89 | 0.88 | 0.88 | 0.90 |
| $\sigma_2$ | 2 | 1.9982 | 0.0624 | -0.0018 | 0.0039 | 0.94 | 0.96 | 0.96 | 0.95 | 0.96 |
| $\theta$ | 5 | 5.4282 | 0.5236 | 0.4282 | 0.4547 | 1.00 | 0.82 | 0.48 | 0.78 | 0.86 |

Notes: T.V. = True Value, S.D. = Standard Deviation, MSE = Mean Squared Error, and C.P. = Coverage Probability

**Table 5:** Estimation results of bivariate Clayton copula-based SUR Tobit with normal margins ($\theta = 10$).

| | | | | | | C.P. | | | | |
|---|---|---|---|---|---|---|---|---|---|---|
| | T.V. | Mean | S.D. | Bias | MSE | Standard Normal (Jackknife) | Standard Normal (Bootstrap) | Percentile | BCa | Basic |
| $\beta_{01}$ | 2 | 2.0056 | 0.0326 | 0.0056 | 0.0011 | 0.89 | 0.89 | 0.88 | 0.89 | 0.88 |
| $\beta_{11}$ | 1 | 0.9950 | 0.0308 | -0.0050 | 0.0010 | 0.92 | 0.92 | 0.92 | 0.92 | 0.91 |
| $\sigma_1$ | 1 | 0.9977 | 0.0218 | -0.0023 | 0.0005 | 0.94 | 0.94 | 0.93 | 0.93 | 0.95 |
| $\beta_{02}$ | 4 | 4.0135 | 0.0767 | 0.0135 | 0.0061 | 0.86 | 0.89 | 0.90 | 0.90 | 0.88 |
| $\beta_{12}$ | -0.5 | -0.5042 | 0.0303 | -0.0042 | 0.0009 | 0.91 | 0.89 | 0.89 | 0.88 | 0.90 |
| $\sigma_2$ | 2 | 1.9966 | 0.0384 | -0.0034 | 0.0015 | 0.96 | 0.98 | 0.96 | 0.98 | 0.98 |
| $\theta$ | 10 | 9.8602 | 0.4869 | -0.1398 | 0.2566 | 1.00 | 0.90 | 0.90 | 0.91 | 0.89 |
| $\beta_{01}$ | 1.5 | 1.5066 | 0.0329 | 0.0066 | 0.0011 | 0.91 | 0.91 | 0.89 | 0.89 | 0.90 |
| $\beta_{11}$ | 1 | 0.9935 | 0.0310 | -0.0065 | 0.0010 | 0.93 | 0.94 | 0.93 | 0.95 | 0.95 |
| $\sigma_1$ | 1 | 0.9968 | 0.0231 | -0.0032 | 0.0005 | 0.93 | 0.93 | 0.93 | 0.93 | 0.93 |
| $\beta_{02}$ | 3 | 3.0138 | 0.0777 | 0.0138 | 0.0062 | 0.84 | 0.86 | 0.86 | 0.87 | 0.88 |
| $\beta_{12}$ | -0.5 | -0.5042 | 0.0312 | -0.0042 | 0.0010 | 0.91 | 0.93 | 0.92 | 0.91 | 0.92 |
| $\sigma_2$ | 2 | 1.9961 | 0.0401 | -0.0039 | 0.0016 | 0.97 | 0.98 | 0.98 | 0.98 | 0.98 |
| $\theta$ | 10 | 9.8763 | 0.5252 | -0.1237 | 0.2911 | 1.00 | 0.89 | 0.89 | 0.89 | 0.89 |
| $\beta_{01}$ | 1 | 1.0064 | 0.0346 | 0.0064 | 0.0012 | 0.91 | 0.91 | 0.90 | 0.90 | 0.90 |
| $\beta_{11}$ | 1 | 0.9960 | 0.0335 | -0.0040 | 0.0011 | 0.91 | 0.91 | 0.91 | 0.92 | 0.91 |
| $\sigma_1$ | 1 | 0.9979 | 0.0275 | -0.0021 | 0.0008 | 0.89 | 0.90 | 0.90 | 0.86 | 0.89 |
| $\beta_{02}$ | 2 | 2.0156 | 0.0768 | 0.0156 | 0.0061 | 0.89 | 0.88 | 0.89 | 0.89 | 0.91 |
| $\beta_{12}$ | -0.5 | -0.5039 | 0.0327 | -0.0039 | 0.0011 | 0.92 | 0.92 | 0.91 | 0.92 | 0.90 |
| $\sigma_2$ | 2 | 1.9983 | 0.0406 | -0.0017 | 0.0016 | 0.98 | 0.98 | 0.98 | 0.98 | 0.97 |
| $\theta$ | 10 | 9.9423 | 0.5468 | -0.0577 | 0.2993 | 1.00 | 0.94 | 0.92 | 0.91 | 0.92 |
| $\beta_{01}$ | 0.5 | 0.5037 | 0.0425 | 0.0037 | 0.0018 | 0.85 | 0.85 | 0.86 | 0.85 | 0.84 |
| $\beta_{11}$ | 1 | 0.9968 | 0.0403 | -0.0032 | 0.0016 | 0.91 | 0.93 | 0.91 | 0.91 | 0.91 |
| $\sigma_1$ | 1 | 0.9988 | 0.0318 | -0.0012 | 0.0010 | 0.88 | 0.89 | 0.88 | 0.87 | 0.89 |
| $\beta_{02}$ | 1 | 1.0135 | 0.0837 | 0.0135 | 0.0072 | 0.86 | 0.84 | 0.85 | 0.85 | 0.84 |
| $\beta_{12}$ | -0.5 | -0.5058 | 0.0358 | -0.0058 | 0.0013 | 0.92 | 0.93 | 0.93 | 0.93 | 0.93 |
| $\sigma_2$ | 2 | 1.9968 | 0.0538 | -0.0032 | 0.0029 | 0.96 | 0.97 | 0.95 | 0.96 | 0.97 |
| $\theta$ | 10 | 9.9667 | 0.7171 | -0.0333 | 0.5153 | 1.00 | 0.86 | 0.87 | 0.87 | 0.84 |
| $\beta_{01}$ | 0.1 | 0.1007 | 0.0488 | 0.0007 | 0.0024 | 0.87 | 0.86 | 0.85 | 0.86 | 0.85 |
| $\beta_{11}$ | 1 | 0.9996 | 0.0460 | -0.0004 | 0.0021 | 0.88 | 0.87 | 0.86 | 0.85 | 0.86 |
| $\sigma_1$ | 1 | 0.9999 | 0.0337 | -0.0001 | 0.0011 | 0.87 | 0.87 | 0.85 | 0.86 | 0.85 |
| $\beta_{02}$ | 0.25 | 0.2682 | 0.0830 | 0.0182 | 0.0072 | 0.90 | 0.90 | 0.89 | 0.88 | 0.89 |
| $\beta_{12}$ | -0.5 | -0.5038 | 0.0378 | -0.0038 | 0.0014 | 0.91 | 0.91 | 0.89 | 0.89 | 0.89 |
| $\sigma_2$ | 2 | 1.9921 | 0.0561 | -0.0079 | 0.0032 | 0.96 | 0.96 | 0.95 | 0.95 | 0.95 |
| $\theta$ | 10 | 10.1430 | 0.8406 | 0.1430 | 0.7200 | 1.00 | 0.88 | 0.87 | 0.83 | 0.85 |

Notes: T.V. = True Value, S.D. = Standard Deviation, MSE = Mean Squared Error, and C.P. = Coverage Probability

**Table 6:** Variable definitions and sample statistics ($n = 500$).

| Variable | Definition | Mean | S.D. |
|---|---|---|---|
| Dependent variables: amount consumed (100 g) | | | |
| Ready-to-eat breakfast cereals | Quantity of ready-to-eat cereals consumed | 0.438 | 0.318 |
| | Among the consuming ($n = 450$; 90% of sample) | 0.487 | 0.297 |
| Fluid milk | Quantity of fluid milk consumed | 2.585 | 2.421 |
| | Among the consuming ($n = 457$; 91.4%) | 2.828 | 2.392 |
| Binary explanatory variables (yes = 1; no = 0) | | | |
| Male | Gender is male | 0.632 | |
| Age 20-30 | Age is 20-30 | 0.114 | |
| Age 31-40 | Age is 31-40 | 0.146 | |
| Age 41-50 | Age is 41-50 | 0.164 | |
| Age 51-60 | Age is 51-60 | 0.166 | |
| Age > 60 | Age > 60 (reference) | 0.410 | |

*Source*: Compiled from the Continuing Survey of Food Intakes by Individuals, U.S. Department of Agriculture, 1994-1996.

Note: S.D. = Standard Deviation

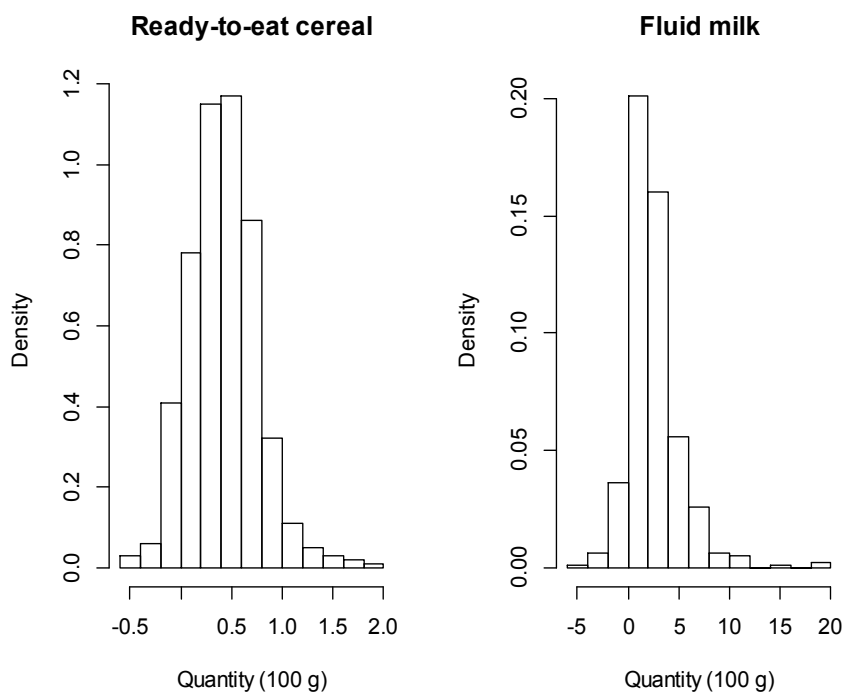

**Figure 2:** Distributions of the ready-to-eat breakfast cereals (left panel) and fluid milk (right panel) consumption.

Table 7 shows the copula and marginal parameter estimates, together with the 90% confidence intervals obtained through the resampling methods described in Section 3.2. It is clearly observed that males consume more RTEC and milk than females. Ganji and Kafai (2004) found similar effect of gender on milk consumption. Compared to those over 60 years of age, individuals age 20-30 consume more fluid milk. The effects of age are slightly different on RTEC, with individuals age 20-30, 31-40 (according to the 90% percentile confidence interval) and 41-50 (according to the 90% bootstrap-based confidence intervals) consuming more than their older counterparts (age > 60). The estimate of the dependence parameter of the Clayton copula ($\hat{\theta} = 1.3284$) and its 90% bootstrap-based confidence intervals show us that the relationship between RTEC and fluid milk is positive (the estimated Kendall's tau is $\hat{\tau} = 0.3991$) and significant at the 10% level (none of the 90% bootstrap-based confidence intervals for this parameter contains the zero value), justifying joint estimation of the censored equations to improve statistical efficiency. Finally, the estimated coefficient of tail dependence for Clayton copula ($\hat{\chi} = 0.5934$), obtained from $2^{-1/\hat{\theta}}$, shows the positive dependence at the lower tail.

## 6. Final Remarks

We extend the analysis of the bivariate SUR Tobit model by modeling its dependence structure through the Clayton copula. The ability to model the tail dependence, especially the lower tail where some data are censored, as well as the flexibility in coupling different marginal distributions, are some of the attractive features of the Clayton copula. We employ the data augmentation technique to generate the unobserved values and proceed the model implementation through the proposed MIFM method. We also propose the usage of resampling methods for obtaining confidence intervals for the model parameters. In the simulation study, we assess the performance of our proposed model and methods through the bivariate Clayton copula-based SUR Tobit model with normal errors, where we obtain the satisfactory results under some specific conditions (high degree of dependence or low to moderate degree of dependence and low to moderate percent censoring in the margins). We then show the applicability of the model and methods to a real dataset of consumption of ready-to-eat breakfast cereals and fluid milk by U.S. individuals. The observed (and significant) positive dependence and the marginal densities (histograms) indicate that modeling this dataset through the bivariate Clayton copula-based SUR Tobit model with normal errors is sufficient. Even though it is rare to analysis the SUR Tobit with more than two dimensions, our proposed model and methods can be straightforwardly applied to high dimensional SUR Tobit models.

**Table 7:** Estimation results of bivariate Clayton copula-based SUR Tobit model with normal margins for ready-to-eat breakfast cereals and fluid milk consumption in the U.S. in 1994-1996.

| Ready-to-eat cereals | Estimate | 90% Confidence Intervals | | | | |
|---|---|---|---|---|---|---|
| | | Standard Normal (Jackknife) | Standard Normal (Bootstrap) | Percentile | BCa | Basic |
| Intercept | 0.3124 | [0.2651; 0.3596] | [0.2579; 0.3668] | [0.2587; 0.3674] | [0.2609; 0.3700] | [0.2573; 0.3661] |
| Male | 0.1024 | [0.0510; 0.1538] | [0.0497; 0.1551] | [0.0510; 0.1565] | [0.0492; 0.1540] | [0.0482; 0.1537] |
| Age 20-30 | 0.1387 | [0.0645; 0.2129] | [0.0513; 0.2262] | [0.0512; 0.2285] | [0.0551; 0.2318] | [0.0490; 0.2262] |
| Age 31-40 | 0.0770 | [-0.0021; 0.1561] | [-0.0016; 0.1555] | [0.0017; 0.1641] | [-0.0022; 0.1602] | [-0.0101; 0.1522] |
| Age 41-50 | 0.0755 | [-0.0061; 0.1570] | [0.0025; 0.1484] | [0.0031; 0.1453] | [0.0058; 0.1482] | [0.0056; 0.1478] |
| Age 51-60 | 0.0320 | [-0.0453; 0.1094] | [-0.0369; 0.1010] | [-0.0352; 0.1044] | [-0.0415; 0.0992] | [-0.0403; 0.0993] |
| S.D. ($\sigma_1$) | 0.3398 | [0.3135; 0.3661] | [0.3204; 0.3592] | [0.3189; 0.3577] | [0.3242; 0.3644] | [0.3219; 0.3607] |
| Fluid milk | Estimate | 90% Confidence Intervals | | | | |
| | | Standard Normal (Jackknife) | Standard Normal (Bootstrap) | Percentile | BCa | Basic |
| Intercept | 1.8777 | [1.5097; 2.2457] | [1.4663; 2.2891] | [1.4653; 2.2668] | [1.4791; 2.2924] | [1.4886; 2.2902] |
| Male | 0.7544 | [0.3714; 1.1374] | [0.3631; 1.1457] | [0.3543; 1.1397] | [0.3325; 1.1103] | [0.3691; 1.1544] |
| Age 20-30 | 0.7794 | [0.1379; 1.4208] | [0.1305; 1.4282] | [0.1389; 1.4160] | [0.1872; 1.4729] | [0.1427; 1.4198] |
| Age 31-40 | 0.2813 | [-0.3835; 0.9461] | [-0.2991; 0.8617] | [-0.2980; 0.8219] | [-0.2566; 0.8714] | [-0.2593; 0.8607] |
| Age 41-50 | 0.2721 | [-0.3706; 0.9147] | [-0.2911; 0.8352] | [-0.2887; 0.8285] | [-0.2824; 0.8312] | [-0.2843; 0.8328] |
| Age 51-60 | -0.3758 | [-0.8333; 0.0816] | [-0.9236; 0.1719] | [-0.8900; 0.1930] | [-0.9351; 0.1405] | [-0.9447; 0.1384] |
| S.D. ($\sigma_2$) | 2.5518 | [2.2138; 2.8899] | [2.3946; 2.7090] | [2.3817; 2.6924] | [2.4299; 2.7802] | [2.4113; 2.7220] |
| $\theta$ | 1.3284 | [-0.5479; 3.2046] | [1.0748; 1.5819] | [1.1342; 1.6398] | [1.0534; 1.5267] | [1.0169; 1.5225] |
| Loglik | -1305.562 | | | | | |

**Acknowledgments**

The research is partially funded by the Brazilian organizations CNPq and FAPESP.